\documentclass[%
 reprint,
 amsmath,amssymb,
 aps,
prl,
twocolumn,
floatfix,
]{revtex4-1}

\usepackage[skip=0pt, indent=10pt]{parskip}

\usepackage{wrapfig}
\usepackage{graphicx}
\usepackage{dcolumn}
\usepackage{bm}

\usepackage{xcolor}
\begin{document}


\title{Precision Measurement of Sub-Continuum Gas Conduction \\ within Micro-Confinements}

\author{Greg I. Acosta}
\author{Malachi Hood}
\author{Mohammad Ghashami}%
 \email{mghashami2@unl.edu}
\affiliation{%
Mechanical and Materials Engineering Department, University of \\
Nebraska-Lincoln, Lincoln, Nebraska 68588, USA
}%





\begin{abstract}
Sub-continuum gas conduction is an essentially important phenomenon in disparate fields of applications ranging from aerospace vehicles to biomedical sensors, and has been the focus of many computational studies over the past decades. These studies predicted that the energy exchange mechanisms are driven by gas-surface interactions, strongly dependent on the gas and surface characteristics.
Despite its fundamental and practical importance, thermal transport via gas conduction at non-continuum regimes mostly remains experimentally unverified.
Here, we report precision measurements of sub-continuum gas conduction within parallel micro-cavities and elucidate its dependence on the gas and surface characteristics. More importantly, we demonstrate a systematic approach for extracting the energy accommodation coefficient (EAC), which is necessary to establish gas-surface scattering kernels or develop diffusive-specular solutions to the Boltzmann transport equation. EACs are also required for calculating the temperature jump coefficient in near-continuum conditions to solve classical hydrodynamical equations. For the first time, we show a correction to the kinetic theory in the transition to near-continuum regimes (particularly for non-monatomic gases) by extracting a physical parameter representing the intermolecular collisions within the Knudsen layer. Our results agree well with the kinetic theory predictions and are expected to inform the development of technologies such as thermal switches, gas sensors, and light-driven actuators.

\end{abstract}

\maketitle


Conductive heat transfer through rarefied gases is a classical and fundamentally important problem in many engineering applications, including thermal insulation of spacecrafts \cite{ uyanna_thermal_2020,xiong_nanoporous_2021,meti_overview_2022}, gas sensors \cite{bandodkar_wearable_2016,baron_amperometric_2017,das_reviewnon-invasive_2020}, micro gas chromatography \cite{qin_fully_2016,leary_deploying_2019}, heat pumps \cite{gupta_thermal_2011,an_monolithic_2015,kugimoto_novel_2018, kugimoto_design_2019, wang2020knudsen}, combustors \cite{xu_concentration_2005,ju_microscale_2011, yang_effect_2019, banerjee_developments_2021}, optical/photophoretic actuators \cite{schmidt2012reconfigurable,kara2015nanofluidics,pennetta_tapered_2016,lu2017light,azadi2021controlled}, and nano-electromechanical systems \cite{passian2003thermal,gotsmann2005experimental,pikus_characterization_2019,vo_measurement_2019,zhu2010origin,defoort2014slippage,gazizulin2018surface}. In such systems, the length scale ($L$) is usually smaller than the molecular mean free path of the gas ($\lambda$), leading to large Knudsen numbers ($\textit{Kn}={\lambda}/{L}$). At $\textit{Kn}>$1, the gas is in a nonequilibrium state which makes the thermal transport process dictated by the gas-surface interactions (GSIs) \cite{mccoy_transport_1974}. Despite the critical implications of rarefied gas conduction, the experimental study of such transport phenomenon has remained very limited \cite{braun_heat_1976,trott_experimental_2011,yamaguchi_investigation_2012,grau2016method,yamaguchi_measurement_2019}. 
Understanding GSIs is inherently complex due to a vast interconnected parameter space associated with gas type, surface material and morphology, surface and gas temperatures, and adsorption susceptibility \cite{yang2022gas,wu2016impact}. To represent the net effect of GSIs, the energy accommodation coefficient (EAC) is commonly used, $\alpha = \Delta E_i/\Delta E_{max}$, where $\Delta E_i$ is the energy that incident gas molecules gain after colliding with the surface, and $\Delta E_{max}$ is the maximum energy attainable from the GSI \cite{goodman2012dynamics}.
EACs are of significant importance in the kinetic modeling of heat transfer problems using Boltzmann Transport Equation (BTE) by providing boundary conditions in the form of temperature jump coefficient for near-continuum conditions \cite{loyalka_temperaturejump_1978,lockerby2004velocity,nguyen2020variational}, or obtaining scattering kernels for GSIs \cite{cercignani1972scattering,wang2021establishing,liao2018prediction}.

Obtaining EACs has been the subject of theoretical \cite{goodman_theory_1965,altman_energy_2020}, numerical \cite{chirita_non-equilibrium_1997,mateljevic_accommodation_2009,khatoonabadi_lattice_2022,kammara_systematic_2019,tokunaga_nonequilibrium_2020,nejad2021modeling}, and experimental studies \cite{braun_heat_1976,rebrov_determination_2003,trott_experimental_2011,yamaguchi_investigation_2012,yamaguchi_measurement_2014,yamaguchi_measurement_2019,sharipov2016energy,ganta2011optical}. While the theoretical studies of EAC strongly rely on the available experimental data for verification and improvement, the measurements have been limited to monatomic and a few polyatomic gases (e.g., N$_2$ and CO$_2$) within ideal confinements. 
Most of the EAC measurements were conducted using the concentric cylinder apparatus \cite{kouptsidis1970accommodation,yeh1973heat,thomas1988heat,o1992experimental,chalabi2012experimental}, which, although simple, could not test different materials or surface structures. To circumvent this limitation, experiments with parallel plate configurations were employed \cite{trott_experimental_2011,bayer2015new,grau2016method}. However, establishing a (micrometer) parallel gap distance and accurately measuring heat fluxes at low pressures has proven challenging, impeding its extensive use.

\begin{figure}[t!]
    \centering
    \includegraphics[width=0.85\linewidth]{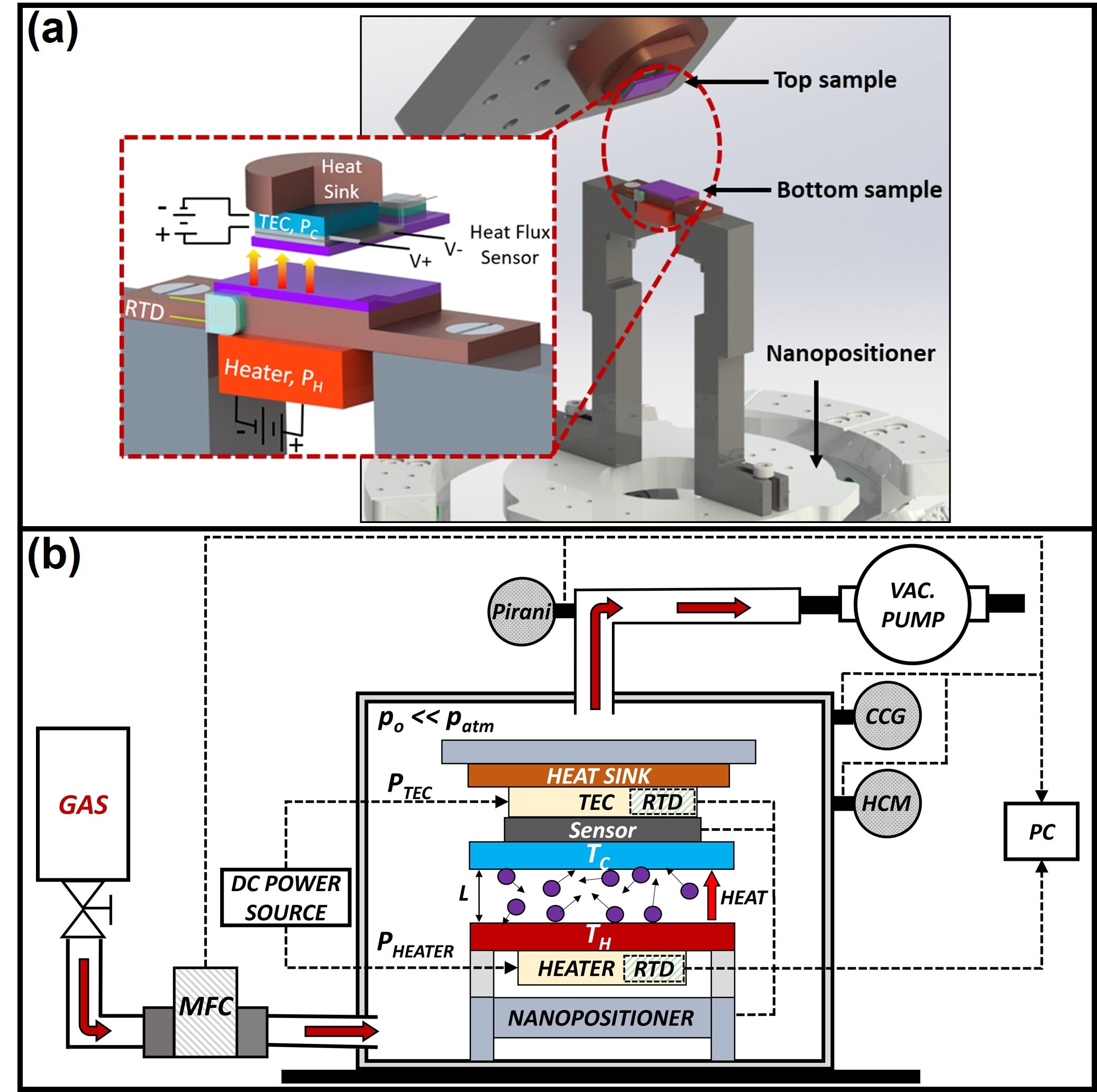}
    \caption{Schematic illustrations of the experimental setup for precision measurement of sub-continuum gas conduction. (a) Developed nanopositioner platform and the inverted breadboard where 10$\times$5~mm$^2$ samples are mounted perpendicularly. The inset shows the arrangement of the sample assemblies. (b) Vacuum chamber housing the setup, equipped with MFC and VAT valve to precisely maintain the gas pressure.}
    \label{fig:my_label2}
\end{figure}

In this letter, we report direct and systematic measurements of sub-continuum gas conduction heat transfer for monatomic, diatomic, and polyatomic gases between planar structures.
We demonstrate accurate extraction of EACs for smooth and functionalized surfaces from the sub-continuum conduction measurements. Furthermore, for the first time, we express a measurement-driven correction factor to the simple kinetic theory in the transition and near-continuum regimes to account for the inter-molecular collisions within the Knudsen layer \cite{gallis_computational_2007,siewert_linearized_2003,loyalka_temperature_1991,loyalka_slip_1990,loyalka_temperaturejump_1978}.
To this end, we have developed a versatile experimental platform (see Fig.~\ref{fig:my_label2}(a)) where two planar samples are mounted with an overlap area of 5$\times$5 mm$^2$. The bottom stage has a heater (Watlow 8$\times$8 mm$^2$) and a resistance temperature detector (RTD) attached to the heat spreader. The top sample is placed on a 4.4$\times$4.4 mm$^2$ heat flux sensor (gSKIN-XM greenTEG) with a response time of 0.7 s and an accuracy of $\pm$3\%. A thermoelectric cooler (TEC) controls the top sample's temperature. RTDs are connected to NI cDAQ-9171 to maintain a steady temperature by feedback controlling the power to the heater and TEC.
To accurately control the relative position of the two surfaces, the top stage is fixed while the bottom stage is placed on a nanopositioner with 1 nm translational resolution in all directions and 1$-\mu$m rotational resolution. Using this nanopositioner enables precise control of the gap distance between the samples, allowing us to reach different thermal transport regimes \cite{ghashami2018precision}. 
As shown in Fig.~\ref{fig:my_label2}(b), the setup is housed in a high-vacuum chamber equipped with a VAT gate valve. A gas supply line is connected to the chamber via a mass-flow-controller (MFC), enabling precise pressure control. 
By adjusting the mass flow rate of the gas from 0$-$5 SCCM, we can maintain steady gas pressures ranging from 0.005$-$0.5 Torr.

For any measurement scenario, the heat transfer between the two samples measured by the heat flux sensor ($Q_{\text{Meas}}$) can consist of two main mechanisms, thermal conduction via gas molecules ($Q_{\text{Gas}}$) and thermal radiation ($Q_{\text{Rad}}$), yielding $Q_{\text{Meas}}=Q_{\text{Gas}}+Q_{\text{Rad}}$.
Thus, as a preliminary step to measuring $Q_{\text{Gas}}$, we must measure the radiative heat transfer, $Q_{\text{Rad}}$, at high-vacuum conditions. $Q_{\text{Rad}}$ consists of the thermal radiation directly exchanged between the samples and the background thermal radiation.
Before any measurements, a parallelism alignment between the two samples is performed to establish a precise gap distance \cite{SI_gasconduction}. After alignment, to measure $Q_{\text{Rad}}$, we fully retract the samples and set $T_\text{H}=50^{\circ}$ and $T_\text{C}=23^{\circ}$ while the pressure is below $10^{-6}$ Torr. Once steady, we measure the heat flux by varying the distance from 750 $\mu$m to 5 $\mu$m in decremental steps while feedback controlling the temperatures.\\
\indent To measure $Q_{\text{Gas}}$, we fully retract the samples to 750~$\mu$m while the temperatures are still fixed. We set the pressure to 0.5 Torr by adjusting the outflow through the VAT valve and controlling the inlet gas flow rate via MFC. Once steady, we record the heat flux and pressure for about 5 minutes. At the same gap, we gradually reduce the gas flow rate to establish lower pressures down to $3.5\times10^{-3}$ Torr. 
The same procedure is repeated for smaller gaps down to 5 $\mu$m. It is noteworthy that for the measurements at lower gaps, we always initialize with a gap of 750 $\mu$m and then approach the samples to the desired distance. This ensures gas particles can occupy the space between the samples, especially at lower pressures. 

\begin{figure}
    \centering
    \includegraphics[width=8.5cm]{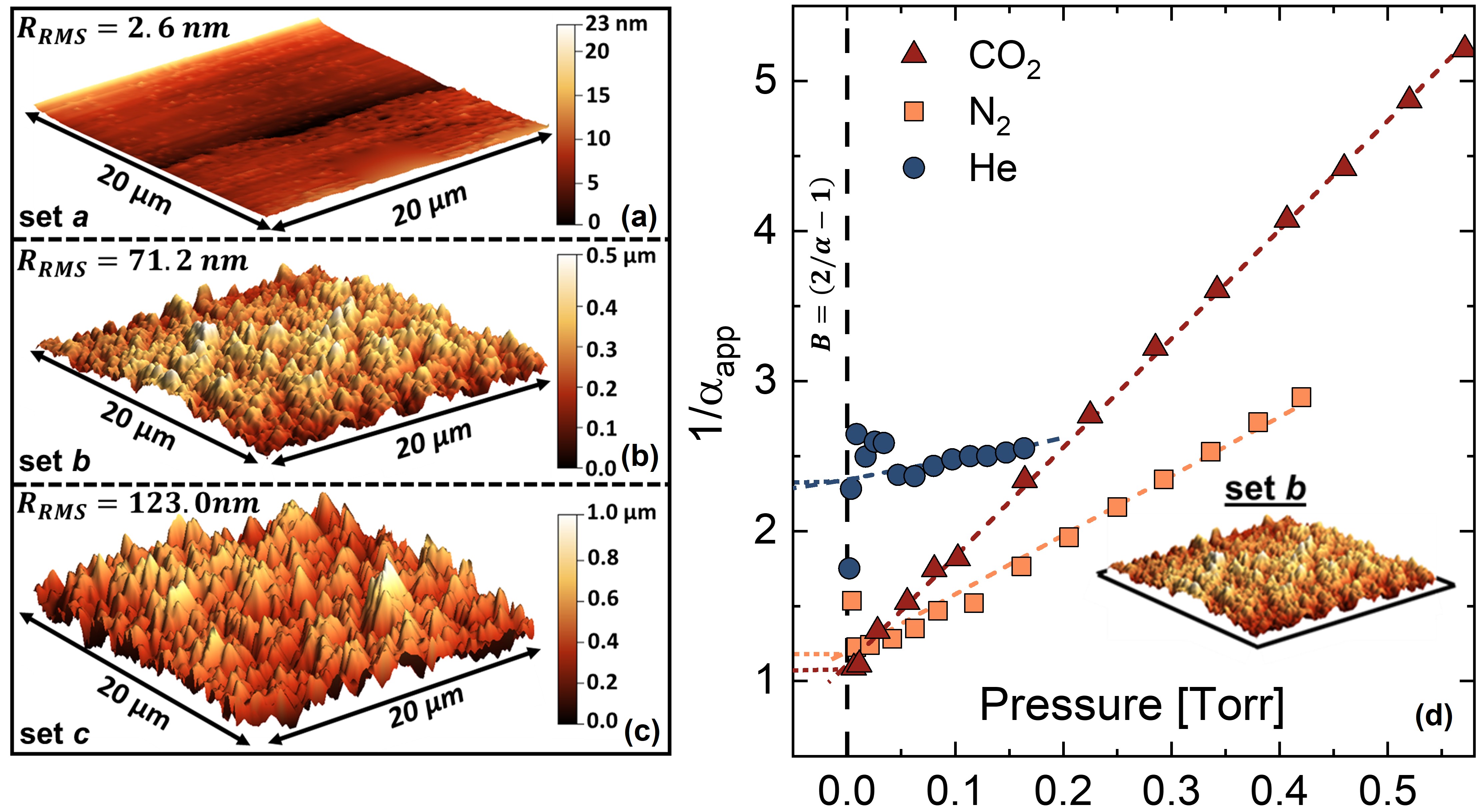}
    \caption{(a)-(c) AFM surface morphology and the measured roughness of the three sets of Si samples. All AFM scans were generated using tapping mode with a Si probe tip, with a resonant frequency of ~260 kHz. (d) EAC extraction from heat flux measurements corresponding to He, N$_2$ and CO$_2$, confined by samples of set $b$ with a gap distance of $L=500~{\mu}m$.}
    \label{fig:my_label3}
\end{figure}

To fully capture the central role of surface characteristics in the energy exchange process through GSI, we used the laser-induced periodic surface structuring (LIPSS) technique to fabricate surface structures in a well-controlled manner. Briefly, LIPSS can generate highly reproducible micro/nanoscale quasi-periodic structures formed due to light-matter interactions between incident ultrashort laser and surface waves that propagate or scatter at the surface of the irradiated material. By modulating the intensity or scanning velocity of the focused laser, one can control the periodicity and the height of the structures ranging from several nanometers to a few micrometers \cite{dusser2010controlled,tsibidis2012dynamics,florian2020surface}. 
In this work, we studied three different sets of samples, all diced out of an N-doped silicon (Si) wafer, with an average resistivity of 1.2 $\Omega\cdot$cm and a crystal orientation of \{111\}. Samples of set $a$ were unmodified to serve as our baseline for comparison to the literature. Samples of sets $b$ and $c$ were fabricated using LIPSS with different raster speeds \cite{acosta2022emissivity, SI_gasconduction}. Fig.~\ref{fig:my_label3} shows the surface morphology of these three sets captured by atomic force microscopy (AFM). As shown, a root-mean-square roughness ($R_{RMS}$) of 2.6 nm was measured for set $a$, while for the LIPSS samples of sets $b$ and $c$, $R_{RMS}$ of 71.2 nm and 123 nm were measured, respectively. After fabrication,  a cleaning protocol was carried out to remove any surface contaminations \cite{wachman1994contributions, SI_gasconduction}. 

The EAC for a particular surface can be determined by measuring the gas conduction heat flux at very low pressures. This technique, known as the Low-Pressure (LP) method \cite{Saxena1989}, requires the measurement of gas conduction at or near the free-molecular regime to minimize the uncertainties due to particle-particle collisions. In that case, EAC can be obtained as the ratio between the measured heat flux ($Q_{\text{Meas}}$) and the theoretical free-molecular heat flux for a fully accommodating surface ($Q_{\text{FM},\alpha=1}$), i.e., ${Q_{\text{Meas}}}/{Q_{\text{FM},\alpha=1}}$  \cite{wachman1962thermal,goodman_thermal_1974,Saxena1989}.
However, measuring heat flux at the deep free-molecular regime might be challenging due to the scarce presence of gas particles, which leads to a weak signal hardly detectable by the heat flux sensor. Therefore, to account for the deviation from purely free-molecular conditions, a different term known as the ``apparent EAC'' is used that can be represented as \cite{cercignani1969rarefied}, 
\begin{equation}
    \alpha_{app}=\frac{Q_{\text{Meas}}}{Q_{\text{FM},\alpha=1}}=\frac{q^*}{1+q^*\left[\left(\frac{1-\alpha_1}{\alpha_1}\right)+\left(\frac{1-\alpha_2}{\alpha_2}\right)\right]}
    \label{app}
\end{equation}
It should be noted that $\alpha_{app}$ is different from the EAC. Here, $q^*=Q_{\alpha=1}/Q_{\text{FM},\alpha=1}$ is a theoretical heat transfer coefficient, where $Q_{\alpha=1}$ is the theoretical gas conduction heat flux at sub-continuum transport regime for a fully accommodating case \cite{thomas1988heat}. 
$\alpha_1$ and $\alpha_2$ are the EACs for the two surfaces, and if identical, $\alpha_1=\alpha_2=\alpha$. In this case, to extract EAC, 
we can rewrite Eq.~\ref{app} in a linear form, ${1}/{\alpha_{app}}=\left({1-q^*}\right)/{q^*}+\left({2}/{\alpha}-1\right)= A(P) + B $, where $A(P)$ varies with pressure and $B$ is the intercept. 
At highly rarefied conditions, $q^*=1$ which yields $\alpha_{app} = \alpha/(2-\alpha)$. Using this factor, the theoretical free-molecular heat flux expression can be given as~\cite{zhang2007nano}, 
\begin{equation}
    Q_{\text{FM}}= \frac{\alpha}{(2-\alpha)}\frac{(\gamma+1)c_vP(T_H-T_C)}{\sqrt{8\pi R T_{FM}}}
    \label{app4}
\end{equation}
where $\gamma$ is the specific heat ratio, $c_v$ is the specific heat capacity at constant volume, $R$ is the specific gas constant, and $T_{\textit{FM}}$ is the effective mean temperature of the gas in the free-molecular regime.

Fig.~\ref{fig:my_label3}(d) shows the linear dependence of $1/\alpha_{app}$ on pressure for He, N$_2$, and CO$_2$ gases with samples of set $b$ while separated by $L=500~\mu m$. A simple Linear Least Square regression is performed to obtain the intercept $B= (2/\alpha - 1)$. The extracted EACs for all surfaces are shown in Table \ref{table:2}. 
\begin{table}[htp]
\centering
\caption{EACs of sample sets for He, N$_2$ and CO$_2$.}
\label{table:2}
\begin{tabular}{||c|c|c| c| c||} 
 \hline
 Sample & $R_{RMS}$ & $\alpha_{\text{He}}$ & $\alpha_{\text{N}_2}$ & $\alpha_{\text{CO}_2}$ \\ [0.5ex] 
 \hline\hline
 set $a$ & 2.6 nm & 0.50 $\pm$ 0.03  & 0.87 $\pm$ 0.04 & 0.97 $\pm$ 0.05\\ 
  set $b$ & 71.2 nm & 0.61 $\pm$ 0.05 & 0.94 $\pm$ 0.06 & 0.97 $\pm$ 0.05 \\
  set $c$ & 123.0 nm & 0.67  $\pm$ 0.02 & 0.99 $\pm$ 0.08 & 0.99 $\pm$ 0.01\\
 [1ex] 
 \hline
\end{tabular}
\end{table}
While the values for He and N$_2$ on smooth Si surfaces agree with the prior experiments \cite{trott_experimental_2011}, there is no literature$-$to our knowledge$-$either on the non-smooth Si surfaces or for the CO$_2$ gas. 
For a given surface, the EAC increases with an increase in the molecular weight and structure of the interacting gas \cite{kouptsidis1970accommodation,goodman2012dynamics}. Expectedly, an increase in surface roughness leads to larger EACs as it causes multiple collisions of the incident gas molecules with the surface. The impact of roughness on EAC enhancement appears to be more prominent for He \cite{hegazy2016thermal}, mainly because of its smaller monatomic structure, which makes it more susceptible to be accommodated by the added roughness \cite{song1987correlation, Demirel1996}. On the contrary, EAC for CO$_2$ shows no change across set $a$ and $b$ due to the inherent nature of the gas molecule to strongly accommodate onto surfaces.

Using the extracted EACs, we can compare the gas conduction measurements with the theoretical predictions from kinetic theory by adopting the expression for the heat flux as \cite{kennard1938kinetic},
\begin{equation}
    Q = \frac{k(T_H-T_C)}{L\left(1+\textit{Kn}~\frac{2-\alpha}{\alpha}~\frac{9\gamma+1}{\gamma-1}\sqrt{\frac{T_{\textit{m,DF}}}{T_{\textit{m,FM}}}}\right)}
    \label{q_th}
\end{equation}
which represents a temperature jump near the surfaces due to the ballistic gas-surface interactions, and a diffusive middle
layer due to particle-particle collisions.
Here, $k$ is the thermal conductivity of the gas \cite{zhang2007nano}, and $T_{\textit{m,DF}}$ and $T_{\textit{m,FM}}$ are the effective mean temperature of the gas at diffusive and free-molecular conditions, respectively. It can be shown that Eq.~\ref{q_th} can also be obtained from the Sherman interpolation, ${1}/{Q} = {1}/{Q_\text{C}}+{1}/{Q_{\text{FM}}}$, where $Q_\text{C} = k(T_H-T_C)/L$ \cite{sherman1963survey}.

\begin{figure}
\centering
{\includegraphics[width=8.5cm]{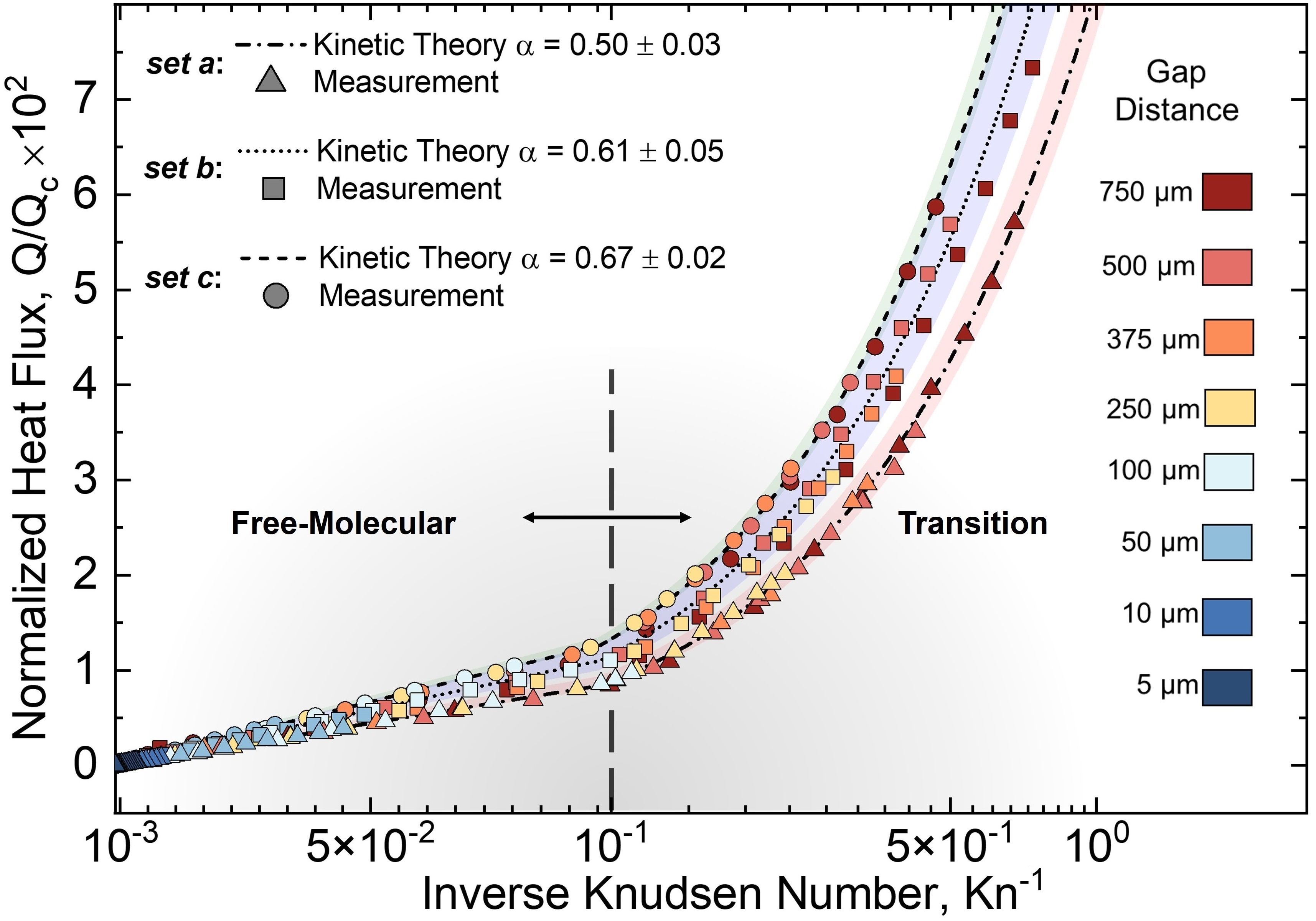}}
\caption{Measured heat flux in the transition and free-molecular regimes for He confined by samples of sets $a$, $b$, and $c$, compared to the kinetic theory predictions using the extracted EACs. The color bands show the effects of EAC uncertainties on the kinetic theory calculations.}
\label{fig:my_label4}
\end{figure}

Fig.~\ref{fig:my_label4} shows the measured gas conduction results for He, corresponding to sets $a$, $b$, and $c$, compared to the theoretical calculations from kinetic theory. The results are normalized with respect to the continuum limit heat flux ($Q_\text{C}$) and plotted as a function of the inverse Knudsen number to fully represent the variation of pressure and gap distance. An excellent agreement is demonstrated between the measurements and the kinetic theory predictions using the extracted EACs, across the free-molecular and transition regimes.
The solution to the heat transfer within the transition regime for monatomic gases under small temperature ratios was previously obtained by solving the BTE under some simplifying assumptions, resulting in Eq.~\ref{q_th} \cite{lees1962kinetic, springer1971heat}. These assumptions allowed the introduction of hard-sphere particles, justified by the simple spherical molecular structure of monatomic gases, which carry only translational kinetic energy. Additionally, assuming a small temperature ratio between the gas and surface will allow linearizing the BTE by representing the velocity distribution of the gas molecules as a simple perturbation from the ideal Maxwell velocity distribution \cite{williams2001review}. Although our measurements were performed at a temperature ratio of approximately 2.0, it has been shown that the EAC of He will not vary with respect to the temperature ratio \cite{kouptsidis1970accommodation}.

\begin{figure*} 
\centering
\makebox[\textwidth]{\includegraphics[width=.8\paperwidth]{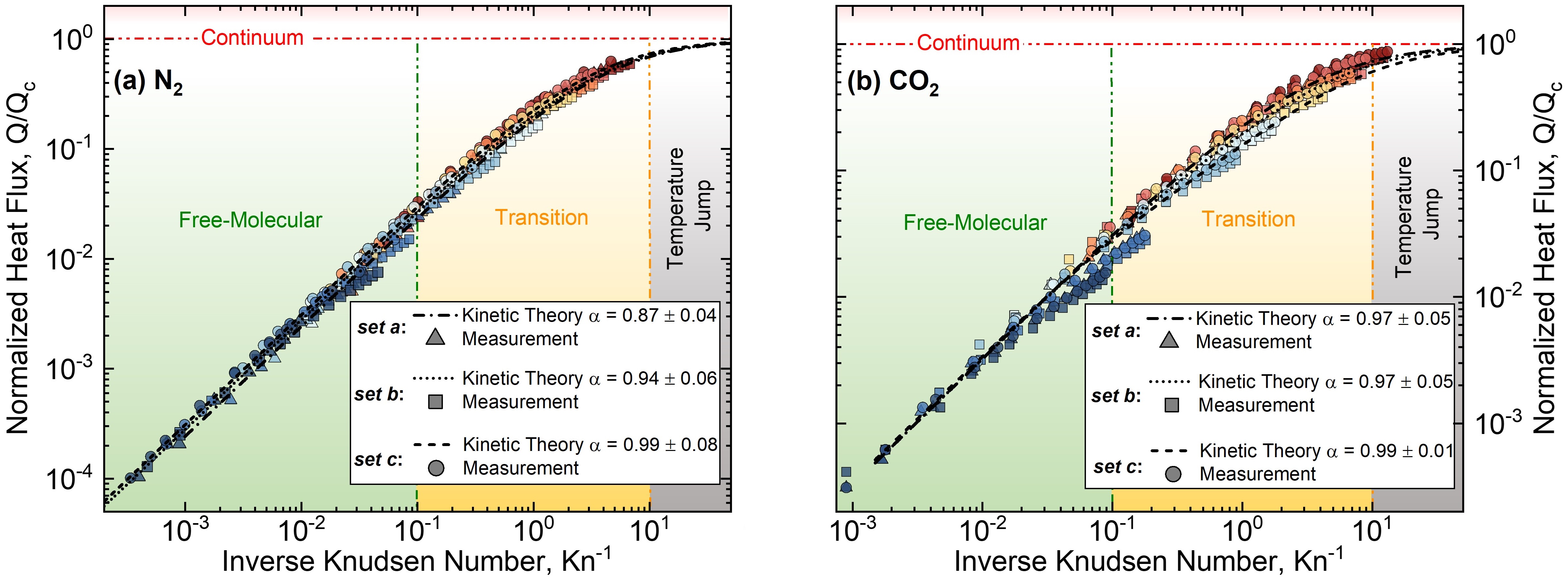}}
\caption{Gas conduction measurements compared to the theoretical predictions from the corrected kinetic theory of Eq.~\ref{q_th_v2} for (a) N$_2$ and (b) CO$_2$ across smooth and functionalized surfaces. The gap distances follow the legend in Fig.~\ref{fig:my_label4}.}
\label{fig:Fig5}
\end{figure*}

Fig.~\ref{fig:Fig5} shows the measurements of sub-continuum gas conduction for N$_2$ and CO$_2$ within the three different sets of surfaces. While most of the measured data for He lies within the free-molecular regime (see Fig.~\ref{fig:my_label4}), it is more evenly distributed for the cases of N$_2$ and CO$_2$. This is attributed to the larger mean free path (due to a smaller molecular diameter) of He compared to N$_2$ and CO$_2$ at the given pressures. 
Notably, at any gap distance, the measured data for N$_2$ and CO$_2$ (particularly at higher pressures) are more dispersed than the He case. This divergence roots in the discrepancy between the gas pressure measured by the transducer and the actual pressure within the two plates. Since the chamber is significantly larger than the measurement system, the local pressure of the gas confined between the two samples is slightly lower than what the pressure transducers are measuring. This deviation grows for gases with larger molecular structures, leading to a miscalculation of the Knudsen number by overestimating the mean free path \cite{SI_gasconduction}.

The measurements show that regardless of the gas type, the heat flux enhances by increasing the surface roughness, as expected from the extracted EACs in Table \ref{table:2}.
Comparing the results of He (as a monatomic gas) with the more complex N$_2$ and CO$_2$ gases, we observe that the heat flux increases remarkably as the size and molecular weight of the gas increase. A polyatomic gas molecule such as CO$_2$ can contain 9 various energy modes (3 translational, 2 rotational, and 4 vibrational degrees of freedom), making it a better energy carrier than He (with 3 translational modes) and N$_2$ (with 3 translational, 2 rotational, and 1 vibrational mode). It is noted that the contribution of vibrational modes (for both N$_2$ and CO$_2$) is negligible due to their high characteristic vibration temperature.

While the measurements for He exhibited a great match to the kinetic theory predictions, the measured heat fluxes for N$_2$ and CO$_2$ do not agree well with the model represented by Eq.~\ref{q_th} \cite{SI_gasconduction}. As discussed earlier, Eq.~\ref{q_th} was derived under simplifying assumptions for monatomic gases, not applicable to more complex polyatomic gases. Moreover, Eq.~\ref{q_th} does not account for the particle-particle collisions in the transition regime, thus resulting in an over-prediction of heat transfer.
To correct the kinetic theory for polyatomic gases in the transition and near-continuum regimes, we use the modified version of Eq.~\ref{q_th} as below \cite{gallis_computational_2007,siewert_linearized_2003,loyalka_temperature_1991,loyalka_slip_1990,loyalka_temperaturejump_1978},
\begin{equation}
    Q = \frac{k(T_H-T_C)}{L\left[1+\textit{Kn}~\frac{2-\alpha}{\alpha}~\frac{9\gamma+1}{\gamma-1}\sqrt{\frac{T_{\textit{m,DF}}}{T_{\textit{m,FM}}}}\left(1+\frac{c_1\alpha}{1+c_2\textit{Kn}}\right)\right]}
    \label{q_th_v2}
\end{equation}
which incorporates two additional parameters, $c_1$ and $c_2$, where $c_1$ represents the effect of intermolecular collisions within the Knudsen layers, and $c_2$ helps retain the free-molecular conditions for the given theory. 
This correction was introduced from the solution to the temperature-jump problem for the linearized BTE, in which variational methods or discrete-ordinate methods were employed to obtain accurate solutions for the case of monatomic gases, by considering different interaction potentials (e.g., Maxwell, hard-spheres, Lennard-Jones, and n(r)-6) \cite{loyalka_slip_1990,loyalka_temperature_1991,siewert_linearized_2003}.
These coefficients can also be obtained from DSMC simulations using the above theoretical solution \cite{gallis_computational_2007}. Nevertheless, these coefficients have never been experimentally verified or extracted from sub-continuum gas conduction measurements. 
To obtain $c_1$ and $c_2$, we employ a non-linear regression, where the dependent variable is the natural logarithmic of the heat flux data, and the independent variables are the extracted EACs and the natural logarithmic of the Knudsen number \cite{gallis_computational_2007, SI_gasconduction}. The obtained $c_1$ coefficients are 0.116 for He, 0.148 for N$_2$, and 0.863 for CO$_2$, while the coefficient $c_2$ was fixed to 0.599 for all gases. The results for the corrected kinetic theory using Eq.~\ref{q_th_v2} are shown in Fig.~\ref{fig:Fig5}, demonstrating good agreement with the measurements for both N$_2$ and CO$_2$. Although the correction for He was not necessary, we expect that if the measurements were conducted near the continuum limit, deviations from the kinetic theory of Eq.~\ref{q_th} would be observed \cite{loyalka_temperaturejump_1978,loyalka_slip_1990,loyalka_temperature_1991,siewert_linearized_2003,gallis_computational_2007}. The correction coefficients reported here are experimentally determined for the first time$-$to our knowledge$-$and are of significant importance for calculating temperature jump coefficient in BTE for any gas-solid system \cite{hattori2018slip, candler2019rate}.

To conclude, we performed systematic measurements to provide an unprecedented experimental demonstration of sub-continuum gas conduction beyond monatomic gases across smooth and laser-functionalized Si surfaces. We experimentally extracted EACs to characterize the gas-surface energy interactions accurately.
Further, we showed the deviation of gas conduction measurements in the transition regime from the Sherman-Lee formula due to the strong impact of the Knudsen layer on the transport mechanism. We addressed this by using the corrected closed-form expression with coefficients derived from our measurements. The findings can shed light on the fundamental understanding of intermolecular potentials required to accurately represent particle-particle collisions in complex gas-surface problems.

\begin{acknowledgments}
This work has been supported by the NASA Nebraska Space Grant Fellowship 4403071026409 and the NASA-EPSCoR Mini-Grant 4403071026385. The research was performed in part in the Nebraska Nanoscale Facility: National Nanotechnology Coordinated Infrastructure and the Nebraska Center for Materials and Nanoscience
(and/or NERCF), which are supported by the National Science Foundation under Award ECCS: 2025298, and the Nebraska Research Initiative. We would also like to thank Dr. Craig Zuhlke and Andrew Reicks for their help with sample fabrication and Dr. Abdelghani Laraoui for helping with AFM characterization.
\end{acknowledgments}

\appendix

\bibliography{apssamp}
\end{document}